\documentclass[conference,10pt,letter]{IEEEtran}
\IEEEoverridecommandlockouts
\def\BibTeX{{\rm B\kern-.05em{\sc i\kern-.025em b}\kern-.08em
    T\kern-.1667em\lower.7ex\hbox{E}\kern-.125emX}}

\pdfoutput=1 

\usepackage[utf8]{inputenc}
\usepackage[most]{tcolorbox}

\usepackage{booktabs}

\usepackage{graphicx}
\usepackage{multirow}
\usepackage[noadjust]{cite} 
\usepackage{setspace}
\usepackage{amsmath}
\usepackage{amssymb} 
\usepackage{amsfonts} %
\usepackage{calligra} %
\usepackage{mathtools}
\usepackage{amsthm} 
\usepackage[hidelinks,bookmarks=false]{hyperref}
\usepackage{dblfloatfix} 
\usepackage{array}
\usepackage{enumitem}

\usepackage{pgfplots}
\usepackage{pgfplotstable}
\usepgfplotslibrary{statistics}

\makeatletter
\let\MYcaption\@makecaption
\makeatother
\usepackage[font=footnotesize]{subcaption}
\makeatletter
\let\@makecaption\MYcaption
\makeatother

\usepackage{comment}
\usepackage{listings}

\usepackage{circledsteps}

\usetikzlibrary{patterns}

\usepackage{blindtext}

\usepackage{acronym}

\usepackage{xcolor, colortbl}

\usepackage{float}
\usepackage{textcomp}

\usepackage{algorithm}
\usepackage{algorithmic}

\usepackage[all=normal, floats=tight,indent=tight]{savetrees}
\usepackage{pifont}
\usepackage{framed}
\usepackage{soul}
\usepackage{multirow}
\usepackage{array}
\newcolumntype{L}[1]{>{\raggedright\let\newline\\\arraybackslash\hspace{0pt}}m{#1}}
\newcolumntype{C}[1]{>{\centering\let\newline\\\arraybackslash\hspace{0pt}}m{#1}}
\newcolumntype{R}[1]{>{\raggedleft\let\newline\\\arraybackslash\hspace{0pt}}m{#1}}
\newcolumntype{H}{>{\collectcell\lstinline}l<{\endcollectcell}}
\usepackage{url}

\input{section/preamble/preamblelistingsconf}
\acrodef{CPS}{Cyber-Physical System}
\acrodef{IoT}{Internet of Things}
\acrodef{HDL}{Hardware Description Language}
\acrodef{CAD}{Computer-Aided Design}
\acrodef{EDA}{Electronic Design Automation}
\acrodef{HPC}{High-Performance Computing}
\acrodef{DL}{deep learning}
\acrodef{ML}{machine learning}
\acrodef{NLP}{natural language processing}
\acrodef{IC}{Integrated Circuit}
\acrodef{CWE}[CWE]{Common Weakness Enumeration}
\acrodef{CVE}[CVE]{Common Vulnerabilities and Exposures}
\acrodef{LLM}[LLM]{large language model}
\acrodef{NMT}[NMT]{neural machine translation}
\newcommand{\ignore}[1]{{}}

\newcommand{\squishlist}{
	\begin{list}{$\bullet$}
		{ \setlength{\itemsep}{0pt}
			\setlength{\parsep}{1pt}
			\setlength{\topsep}{1pt}
			\setlength{\partopsep}{0pt}
			\setlength{\leftmargin}{0.9em}
			\setlength{\labelwidth}{1.5em}
			\setlength{\labelsep}{0.4em} } }
	\newcommand{\squishend}{
	\end{list}  }

\definecolor{graphFirst}{RGB}{2,136,209} 
\definecolor{graphSecond}{RGB}{211,47,47} 
\definecolor{graphThird}{RGB}{245,124,0} 
\definecolor{graphFourth}{RGB}{56,142,60} 
\definecolor{graphFifth}{RGB}{81,45,168} 
\definecolor{graphSixth}{RGB}{69,90,100} 
\definecolor{graphSeventh}{RGB}{251,192,45} 
\definecolor{backgroundSecond}{RGB}{239,154,154} 
\definecolor{backgroundThird}{RGB}{255,204,128} 
\definecolor{backgroundFourth}{RGB}{165,214,167} 
\definecolor{backgroundFifth}{RGB}{179,157,219} 
\definecolor{backgroundSixth}{RGB}{176,190,197} 
\definecolor{backgroundSeventh}{RGB}{255,245,157} 

\begin{document}


\title{
Security and Interpretability in Automotive Systems
}

\author{%
\IEEEauthorblockN{Shailja Thakur}

\IEEEauthorblockA{{University of Waterloo}\\ \textit{s7thakur@uwaterloo.ca, +1 917-283-1342}}
}

\maketitle

\thispagestyle{plain}
\pagestyle{plain}

\begin{abstract}
The lack of any sender authentication mechanism in place makes CAN (Controller Area Network) vulnerable to security threats. For instance, an attacker can impersonate an ECU (Electronic Control Unit) on the bus and send spoofed messages unobtrusively with the identifier of the impersonated ECU.  To address the insecure nature of the system, this thesis demonstrates a sender authentication technique that uses power consumption measurements of the electronic control units (ECUs) and a classification model to determine the transmitting states of the ECUs. The method's evaluation in real-world settings shows that the technique applies in a broad range of operating conditions and achieves good accuracy.

A key challenge of machine learning-based security controls is the potential of false positives. A false-positive alert may induce panic in operators, lead to incorrect reactions, and in the long run cause alarm fatigue. For reliable decision-making in such a circumstance, knowing the cause for unusual model behavior is essential. But, the black-box nature of these models makes them uninterpretable. Therefore, another contribution of this thesis explores explanation techniques for inputs of type image and time series that (1) assign weights to individual inputs based on their sensitivity toward the target class, (2) and quantify the variations in the explanation by reconstructing the sensitive regions of the inputs using a generative model.

 In summary, this thesis (Link: \url{https://uwspace.uwaterloo.ca/handle/10012/18134}) presents methods for addressing the security and interpretability in automotive systems, which can also be applied in other settings where safe, transparent, and reliable decision-making is crucial.

\end{abstract}


\section{Introduction}
\label{sec:intro}

CAN is the most commonly found bus protocol in modern automobile systems today. The two-wire bus protocol helps accomplish sophisticated vehicle services in real-time through complex interactions between hardware components. However, one of the fundamental limitations of CAN is the lack of any sender authentication mechanism in place, which makes CAN vulnerable to security threats. For instance, one of the vehicle functionalities is ADAS (adaptive driver's assistance system) which uses a fusion of sensors connected to the CAN network to help regulate the target vehicle's speed. However, a delay introduced in the vehicle adversarially will cause the system to fail to increase or decrease the speed on time. This unexpected behavior will trigger a cascade of incorrect actions, resulting in a collision. It is also equally likely that the unusual behavior of the system is a result of an error in the system or a fault in the vehicle's automotive components. And to perform reliable decision-making in such an uncertain situation, it is crucial to identify the root cause of the system's behavior. However, due to the complex nature of the system, accountability for the system's unexpected behavior is only partially established. In this dissertation, we propose a technique for authenticating the senders of the messages observed on the CAN bus, followed by a solution for reasoning the predictions of the authentication technique.

\section{Proposed Method}
\label{sec:training}
This section provides an overview of the proposed methods for defending CAN bus against impersonation attack and methods for explaining the decision-making of the proposed approach.
\subsection{Sender Authentication for CAN (Controller Area Network) protocol} 

Given the insecure nature of the CAN bus, several approaches~\cite{scission,viden,cloaking-clock} have been proposed for sender authentication in the CAN protocol. However, those methods rely on authenticating using physical characteristics such as clock skew, voltage signal, and timing information, which can be profiled and imitated by an external device. Therefore, we propose CANOA~\cite{canoa}, a novel technique for sender authentication using power-consumption measurements of ECUs to authenticate the sender of a message. The method exploits the physical characteristics of the transmitting and non-transmitting states of the ECU, which a machine learning model classifies to determine whether the message originated from the purported sender. 

We tested our technique in a lab-prototype setting and real-world setting across different vehicles' for the transferability and robustness of the method. We observe that the approach can be used to detect the presence of compromised and additional devices on the network. We also observe that the evaluation of our approach in a range of real-world settings, such as Sterling Acterra Truck, Heavy-duty vehicles over a long period and under different operating conditions, helps attain a false positive rate of 0.01\%. We also show that the approach applies to different network settings and works with limited computing resources without compromising on accuracy. 

\subsection{Explainability}

One of the challenges of machine learning-based security controls is the likelihood of false alerts, wrongly denoting the presence of an attack, which can induce panic in the vehicle operator and leads to fatal consequences. Understanding why the system erroneously detects attacks is imperative for rational decision-making in such a context. However, the complex nature of these techniques makes it difficult to determine the cause of the system's behavior. A significant body of prior work~\cite{lime,preturbation,grad-cam,saliency-map} attempts to interpret model insights. However, these methods rely on models' parameters and weights, accessible by unwanted tampering with the warranty of the models. However, a solution based on non-intrusive techniques by Petsiuk~et~al.~\cite{rise} depends on random input perturbations, which are computationally intensive and lack consistency. Hence, their usage is limited in the safety-critical domain. 
Therefore, we propose a solution comprising three closely related sub-problems that help to explain the outcomes of the time series-based sender authentication technique. 

The first sub-problem addresses the limitation of the prior work and proposes a non-intrusive, model-agnostic, and computationally fast explanation method. Our work~\cite{explainability} builds upon the previous work by implementing a non-intrusive perturbation-based technique that uses \textit{empirical risk minimization} to optimize a randomly initialized input mask. Because the method works by iteratively retaining and propagating pixels sensitive for mapping to the target class, it converges to an estimate of saliency-map faster than~\cite{rise}. 

We evaluated our saliency map-based explanation approach using two datasets: ImageNet and MS-COCO, and against a set of models: VGG16, ResNet50, and Inception V3. We also compared our approach against other saliency map-based techniques. Experiments show that our method performs better than the existing explainability approaches across most of the tested scenarios.

A limitation of the perturbation-based explanation technique is that it lacks consistency. As a second sub-problem, we address the instability of the explanation and propose a method~\cite{explainability} to generate variations for the salient region of the input for which the model prediction remains unaltered. For generating alternative explanations, we used an image completion technique~\cite{image-inpainting}, which reconstructs the pixels in the salient regions of the input by locating encodings in the latent space that are closest to the encoding of the neighboring pixels. 

Using this approach, we find an exhaustive and contextually similar set of transformations for the pixels in the semantic regions, which are also classified into the target class. The accuracy of the reconstructed input using the saliency map as the mask shows that the approach can find explanations relevant to the classification of input to the target class.

Unlike images, temporal sequences such as power-consumption measurements are less intuitive to humans for interpreting the decision-making process. Therefore, we leverage the proposed model-agnostic and generalized explanation method for images to explain the outcome of the time series-based sender authentication technique. To do that, we refine our perturbation-based explanation method for time series-based models and propose a non-intrusive and class-discriminative explanation technique, TiME (\underline{ti}me series based \underline{m}odel outcome \underline{e}xplanation). The method assigns scores to every input time unit based on their significance toward the target class as an expectation over the weighted random input sub-samples. The weights are the model's confidence in the target classes. The sub-samples are chosen because they are contiguous and windowed segments of the inputs to avoid introducing spurious artifacts in the sub-samples. 

The evaluation of the technique against a wide range of publically available UCI time series datasets and state-of-the-art models shows that the approach learns to focus and discriminate between the relevant inputs for classification to the output class. We also evaluate the method for discrimination between the power consumption measurement segment that results in false positives and true positives. We observed that the technique can highlight the distinct time-points in the input, which are sensitive for the classification to the incorrect class of ECU.

\section{Conclusions and Future Work}

\label{sec:conclusions}
This thesis proposes a novel approach for securing automotive systems and interpreting the decision-making of the machine learning-based classifier used in automotive systems. The proposed techniques can be applied to other avenues where the security and interpretability of the decision-making system are crucial. The sender authentication followed by a model outcome explainability technique alleviates the gap between the decision-making process and the rationale behind the model's decisions.



\bibliographystyle{IEEEtran}
\bibliography{lit/references}

\end{document}